\documentclass[pra,aps,twocolumn,showpacs,superscriptaddress,floatfix,amsmath,amssymb,nofootinbib]{revtex4}
\newcommand{\beq}{\begin{equation}}
\newcommand{\eeq}{\end{equation}}
\newcommand{\beqa}{\begin{eqnarray}}
\newcommand{\eeqa}{\end{eqnarray}}
\def\ra{\rangle}
\def\la{\langle}
\usepackage{graphicx}
\usepackage{bm}

\begin{document}

\title{Doubly resonant ultrachirped pulses}
\author{S. Ib\'a\~{n}ez}
\affiliation{Departamento de Qu\'{\i}mica F\'{\i}sica, Universidad del Pa\'{\i}s Vasco - Euskal Herriko Unibertsitatea,
Apdo. 644, Bilbao, Spain}
\author{A. Peralta Conde}
\affiliation{Departamento de Qu\'{\i}mica F\'{\i}sica, Universidad del Pa\'{\i}s Vasco - Euskal Herriko Unibertsitatea,
Apdo. 644, Bilbao, Spain}
\author{D. Gu\'ery-Odelin}
\affiliation{Laboratoire Collisions Agr\'egats R\'eactivit\'e, CNRS UMR 5589, IRSAMC, Universit\'e Paul Sabatier, 118 Route de Narbonne, 31062 Toulouse CEDEX 4, France}
\author{J. G. Muga}
\affiliation{Departamento de Qu\'{\i}mica F\'{\i}sica, Universidad del Pa\'{\i}s Vasco - Euskal Herriko Unibertsitatea,
Apdo. 644, Bilbao, Spain}

\begin{abstract}
%
Ultrachirped pulses for which the frequency chirp is of the order of the transition frequency of a two-level atom are examined. When the chirp is large enough, the resonance may be crossed twice, for positive and negative quadrature frequencies.  In this scenario the analytic signal and quadrature decompositions of the field into amplitude and phase factors turn out to be quite different.
The corresponding interaction pictures are strictly equivalent, but only as long as approximations are not applied. 
The domain of validity of the formal rotating wave approximation 
is dramatically enhanced using the analytic signal representation. 
%
\end{abstract}
\maketitle
\section{Introduction}
%
%
%

Taking the parameters of a system to extreme values beyond their standard domain 
is a common and fruitful exercise in physics. It frequently discloses different regimes, properties or qualitative changes in the system behavior. Appropriate modifications in the theoretical or experimental treatments or plainly new tools may also be required, and often the new physics found leads to unexpected applications.     
Examples of much current interest in atomic, molecular and optical science are ultracold temperatures, or ultrashort and ultrastrong field-matter interactions. In this paper we want to put forward and examine the limit of ultrachirped pulses. By an ``ultrachirped pulse'' we mean one where the instantaneous frequency change is of the order of the transition frequency of a two-level (actual or artificial) atom. We shall describe in particular linearly chirped pulses that  excite the resonance twice, at positive and negative instantaneous quadrature frequencies. 
A consequence is a strong divergence between the analytic-signal 
and the quadrature-model splitting of the field into amplitude and phase factors, 
which causes differences between the corresponding formal rotating wave approximations.  
      
The rotating wave approximation (RWA) is a widely applied simplification to treat radiation-matter interaction neglecting rapidly oscillating terms. Its validity has been analyzed for different applications  
\cite{Fleming}, and the deviations from exact results are currently of much interest due to the increasing ability to manipulate interaction parameter values in different physical settings, and to produce strong couplings and/or ultrashort pulses. 

Except for monochromatic, constant-intensity fields,   
the choice of amplitude and phase to describe the field is not unique, 
and it affects the definition of the Rabi frequency $\Omega_R(t)$, instantaneous frequency $\omega(t)$,  and detuning $\Delta(t)$. This leads to different interaction pictures  and different accuracies for the corresponding rotating wave approximations (RWA).
We demonstrate that the amplitude-phase partition provided by the analytic signal theory enhances significantly the domain of validity of the RWA with respect to the commonly applied quadrature model partition.  
To make the paper self-contained we shall first review briefly in the following subsections the elements of interaction pictures for time dependent parameters and of analytic signal theory. Section II is devoted to our study case, a linearly ultrachirped Gaussian pulse. 
%
%
%
%
%
\subsection{General formulation of intermediate pictures}
Interaction pictures (IP) or ``representations'' \cite{Messiah}, intermediate between the Schr\"odinger and Heisenberg pictures, may be most generally formulated  in terms of a unitary operator $U_0(t)$ \cite{Messiah}, which defines, from the Schr\"odinger picture state $\psi_S(t)$,
the IP state $\psi_I(t)=U_0^\dagger(t)\psi_S(t)$.
$\psi_I(t)$ evolves according to the dynamical 
equation $i\hbar \partial_t \psi_I(t)=H_I(t)\psi_I(t)$,
where 
\beqa
H_I(t)&=&U_0^\dagger(t) H'(t) U_0(t),
\\
H'(t)&=&H(t)-H_0(t),
\\
H_0(t)&=&=i\hbar\dot{U}_0 U_0^\dagger.  
\eeqa
$H(t)$ is the Schr\"odinger picture Hamiltonian, and 
the dot denotes derivative with respect to $t$. 
In this general formulation the primary object is the unitary operator $U_0(t)$, and $H_0(t)$ is ``derived'' as the Hamiltonian for which $U_0(t)$ is an evolution operator. In many textbooks the emphasis is the opposite: the starting point is a splitting of the Hamiltonian $H(t)=H_0(t)+H'(t)$, and then $U_0(t)$ is defined as the evolution operator of $H_0(t)$ \cite{Cohen_Tannoudji}. Quite frequently $H_0(t)$ is time-independent and  $U_0(t)=e^{-iH_0t/\hbar}$, as in typical applications of time-dependent perturbation theory, but we stress that this is only a particular case and by no means necessary.  
\subsection{Interaction pictures for a two-level atom}
We shall assume a semiclassical description of the 
interaction between a laser
electric field linearly polarized in $x$-direction, $\textbf{E}(t)=E(t)\widehat{x}$,  
and a 2-level-atom. Hereafter we shall refer to $E(t)
=E_0(t)\cos[\theta(t)]$ as ``the field''.
In the  
electric dipole approximation, and for the general case where the transition frequency $\omega_0(t)$ may depend on time, the exact Hamiltonian of the atom in the Schr\"odinger picture is 
\beqa
H_{e}(t)&=&\frac{\hbar \omega_{0}(t)}{2} [|2\rangle \langle2|-|1\rangle \langle1|]
\nonumber\\
&+&\frac{\hbar \Omega_{R}(t)}{2}[(|2\rangle \langle1|+|1\rangle \langle2|) (e^{i\theta(t)}+e^{-i\theta(t)})],
\eeqa
where $\theta(t)=\int_0^t{\omega(t')dt'}$ and $\omega(t)=\dot{\theta}(t)$ is the instantaneous frequency. $\Omega_R=\Omega_R(t)=cE_0(t)$ is the Rabi frequency with $c$ a real constant, 
the component of the dipole moment in the polarization direction divided by $\hbar$. 
Note that the definitions of $\omega(t)$ and $\Omega_R$ are not unique,
as emphasized below. It is usually convenient to write the 
phase as $\theta(t)=\omega_L t+\varphi(t)$, 
taking $\omega_L$ as a constant, the carrier frequency, so that   
$\omega(t)=\omega_L+\dot{\varphi}(t)$. 
   
The laser-adapted interaction picture, also called in this context rotating frame, based on defining
\beq
H_0(t)=H_L(t)=\frac{\hbar\omega(t)}{2} (|2\ra\la 2|-|1\ra\la 1|)
\eeq
and
\beqa
U_0(t)&=&e^{-i\int_0^t{H_L(t') dt'}/\hbar}
\nonumber\\
&=&e^{-i\theta(t)/2} |2\rangle \langle2|+e^{i\theta(t)/2} |1\rangle \langle1|,
\eeqa
produces the IP Hamiltonian\footnote{This was termed, with a different notation ``quasi-interaction picture'' in \cite{PRL}. In fact, according to 
the definition of IP given above, it is a perfectly canonical one.}
\beq   
H_{I}(t)=\frac{\hbar}{2}
\left(\begin{array}{cc} 
-\Delta(t) & (1+e^{-2i\theta(t)})\Omega_{R}(t)\\
(1+e^{2i\theta(t)})\Omega_{R}(t) & \Delta(t)
\end{array} \right),
\label{exact_H}
\eeq
where we use the vector basis
$|1\rangle = \left( \begin{array} {rccl} 1\\ 0 \end{array} \right)$, $|2\rangle = \left( \begin{array} {rccl} 0\\ 1 \end{array} \right)$, and
\beq
\label{laser_detuning}
\Delta(t) \equiv \omega_0(t)-\omega(t)
\eeq
is the detuning between the transition frequency and the instantaneous frequency. If $\omega_0(t)$ does not depend on time and $\omega_0=\omega_L$,
as we shall assume hereafter,
\beq\label{inverse}
\Delta(t)=-\dot{\varphi}.
\eeq 

Applying now a RWA to get rid of the counter-rotating terms we end up with 
\beq   
H_{I,RWA}(t)=\frac{\hbar}{2}
\left(\begin{array}{cc} 
-\Delta(t) & \Omega_{R}(t)
\\
\Omega_{R}(t) & \Delta(t)
\end{array} \right). 
\label{HI1}
\eeq
This is a ``formal'' RWA applied  
blindly at this point, without analyzing the frequency content of the neglected 
phase factors, to be distinguished from a more accurate treatment  in which the phase factor is Fourier analyzed and filtered.  
The term ``RWA'' refers here to this approximation and, as we shall see, this 
crude, formal approach will be valid or not depending on the field
partition chosen.           
%
%
%
%
%
%

%
%
%
%
%
%
%
%
%
%
%
%
%
%
\subsection{Phase and instantaneous frequency of a signal}
We summarize here some relevant elements of signal theory \cite{Cohen}. 
A general real signal can be written in the form
\beq
\label{real_signal}
s(t)=a(t)\cos{\theta(t)},
\eeq
where $a(t)$ is the amplitude and $\theta(t)$ the phase. 
This decomposition of $s(t)$, however, is not unique. In other words,
we may find different amplitude and phase functions $a'$ and $\theta'$
satisfying $s(t)=a'(t)\cos{\theta'(t)}$.    

It is customary to define a corresponding complex signal,
\beq
\label{complex_signal}
Z(t)=A(t)e^{i\Theta(t)},
\eeq
with real part $s(t)$, 
%
\beq
\label{real_imaginary parts}
Z(t)=s(t)+is_i(t),
\eeq
and imaginary part $s_i(t)$ defined in different ways.
For a given complex signal the instantaneous frequency is defined as the derivative of the phase, $\omega(t)=\dot{\Theta}(t)$, so it depends on the imaginary part chosen.   

Moments of the signal may be defined as $\la t^n\ra=\int s^2 t^n dt/\int s^2 dt$, and the variance is $\sigma_t^2=\la t^2\ra-\la t\ra^2$. This extends to complex signals as well 
changing $s^2\to|Z|^2$. 
In frequency space, spectral moments are defined similarly in terms of the spectrum 
\beq
\label{spectrum}
S(\omega)=\frac{1}{\sqrt{2\pi}}\int_{-\infty}^\infty s(t) e^{-i\omega t}dt,   
\eeq
and the bandwidth $\sigma_\omega$ is defined as the root of the variance. 
For a real signal $S(-\omega)=S^*(\omega)$ and $|S(\omega)|^2$ is
symmetric about the origin, so $\la \omega\ra=0$.  
It is also possible to extend the spectrum concept to the complex signal, or to 
parts of it, and break this symmetry.  
\subsubsection{The quadrature signal}
From the form (\ref{real_signal}), it is natural to write the complex signal as
\beq
\label{quadrature_signal}
s_q(t)=a(t)e^{i\theta(t)},
\eeq
which is called the quadrature (model) signal. As Eq. (\ref{real_signal}) is not unique,  
the quadrature signal is not unique either. $S_q(\omega)$ is the corresponding spectrum.  
%
%
%
%
%
\subsubsection{The analytic signal}
The analytic signal is a peculiar complex signal chosen as  
\beq
\label{analytic_signal}
s_a(t)=\frac{2}{\sqrt{2\pi}}\int_0^\infty S(\omega) e^{i\omega t}d\omega.
\eeq
The imaginary part $s_{a,i}(t)$ is the Hilbert transform of the real signal
$s(t)$. 
$s_a(t)$ can also be written in terms of its amplitude $A_a(t)=(s^2+s_{a,i}^2)^{1/2}$ and its polar phase $\theta_a(t)$ as
\beq
\label{analytic_signal_2}
s_a(t)=A_a(t)e^{i\theta_a(t)}.
\eeq
%
One of the advantages of the analytic signal  
is that it puts the low frequencies in the amplitude and the high frequencies in the phase factor $e^{i\theta_a(t)}$ \cite{Cohen}. If the spectrum of the amplitude is of finite support, the support of the spectrum of the phase factor does not overlap and this makes the pair $[A_a(t), \theta_a(t)]$ unique.  
The spectrum of $s_a(t)$ is $S_a(\omega)=S_q(\omega)+S_q^*(-\omega)$ if $\omega>0$ and zero otherwise.

The maximum possible deviation of a quadrature signal from the analytic signal, at time $t$, is given by
\beqa
|s_a(t)-s_q(t)|\leq \frac{2}{\sqrt{2\pi}}\int_{-\infty}^0 |{S}_q(\omega)|d\omega.
\eeqa
Another criterion of closeness is that 
\beqa
\label{E_criterion}
\int |s_a(t)-s_q(t)|^2 dt= 2 \int_{-\infty}^0 |{S}_q(\omega)|^2 d\omega
\eeqa
should be small, i.e., the spectrum of the quadrature signal should be predominantly  
in the positive frequency domain \cite{Cohen}.
\section{RWA with the quadrature and the analytic signals for a 
linearly chirped Gaussian pulse}
We consider now 
an electric field with a Gaussian envelope and a linear chirp,
\beq
\label{field}
E(t)={\cal E}_0 e^{-at^2}\cos{(\omega_Lt+bt^2)},
\eeq
as the real signal, where $E_0(t)={\cal E}_0 e^{-at^2}$ is its amplitude. 
Not to restrict the numerical examples to a particular system or region of the 
spectrum, we shall
from now on use dimensionless variables. We define $\tilde{t}=a^{1/2}t$, $\tilde{\omega}=\omega/a^{1/2}$,
and apply the same scaling to all times and frequencies. Similarly, $\tilde{b}=b/a$.  
From the definition of the dimensionless Rabi frequency, $\tilde{\Omega}_R=\Omega_R/a^{1/2}$, the dimensionless amplitude of the field is $\tilde{E}_0=\tilde{\cal E}_0 e^{-\tilde{t}^2}$, where
$\tilde{\cal E}_0= c{\cal E}_0/a^{1/2}$. Dimensionless Hamiltonians are defined 
as $\tilde{H}=H/(a^{1/2}\hbar)$. To avoid a heavy notation we finally   
drop all the tildes hereafter. The dimensionless version of Eq. (\ref{field}) is thus 
\beq
\label{field2}
E(t)={\cal E}_0 e^{-t^2}\cos{(\omega_Lt+bt^2)},
\eeq
in terms of a carrier or central frequency $\omega_L$, a chirp parameter $b$, 
and maximum  
amplitude ${\cal E}_0$.  
In these units, all pulses have equal duration,
with $\sigma_t=1/2$. 
   
\begin{figure}[h]
\begin{center}
\includegraphics[height=4cm,angle=0]{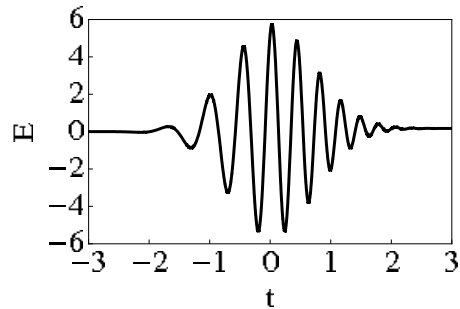}
\end{center}
\caption{Linearly chirped Gaussian pulse. Parameters: $b=2$, ${\cal E}_0=4 \sqrt{2}$, and $\omega_L=10 \sqrt{2}$.}
\end{figure}

The corresponding quadrature model signal is 
\beq
\label{quadrature_field}
E_q(t)={\cal E}_0 e^{-t^2} e^{i(\omega_Lt+bt^2)},
\eeq
with amplitude $E_{0q}(t)={\cal E}_0 e^{-t^2}$, phase $\theta_q(t)=\omega_Lt+bt^2$, and instantaneous frequency $\omega_{q}(t) =\omega_L+2bt$ corresponding to a linear chirp. 
The dimensionless interaction picture Hamiltonian in the rotating wave approximation (\ref{HI1}) becomes 

\beq
H_{I,RWA,q}(t)=\frac{1}{2}
\left(\begin{array}{cc} 
-\Delta_{q}(t) & \Omega_{Rq}(t)\\
\Omega_{Rq}(t) & \Delta_{q}(t)
\end{array} \right),
\label{HI1_q}
\eeq
where $\Delta_{q}(t)=-2bt$ if $\omega_0(t)=\omega_L$. 

The analytic model signal provides instead the complex field
$E_a(t)$ according to Eq. (\ref{analytic_signal}),  
%
%
%
%
%
\beq
\label{analytic_splitting}
E_a(t)=E_{0a}(t)e^{i\theta_a(t)}.
\eeq
The corresponding interaction picture is different from the one using the quadrature model, and the RWA Hamiltonian (\ref{HI1}) becomes
\beq
H_{I,RWA,a}(t)=\frac{1}{2}
\left(\begin{array}{cc} 
-\Delta_{a}(t) & \Omega_{Ra}(t)\\
\Omega_{Ra}(t) & \Delta_{a}(t)
\end{array} \right),
\label{HI1_a}
\eeq
where $\Delta_{a}(t)=\omega_0(t)-\omega_{a}(t)$, with  $\omega_{a}(t)=\dot{\theta}_a(t)$. 

To calculate $E_{a}(t)$ we may use the relation \cite{Cohen}
\beq
E(\omega)=\frac{1}{2}[E_q(\omega)+E_q^*(-\omega)],
\eeq
where $E(\omega)$ is the spectrum of $E(t)$, and 
\beq
E_q(\omega)=\frac{1}{\sqrt{2\pi}}\int_{-\infty}^\infty\!\! E_q(t) e^{-i\omega t}dt
=\frac{{\cal E}_0 e^{-(\omega-\omega_L)^2/4(1-ib)}}{\sqrt{2(1-ib)}}.
\eeq
%
%
%
Then, from Eq. (\ref{analytic_signal}), 
\beqa
\label{Ea_W}
E_a(t)= \frac{{{\cal{E}}_0}}{2}\! \left[e^{-\omega_L^2/4(1-ib)} w(z)\!+\! e^{-\omega_L^2/4(1+ib)} w(z^{\ast})\right]\!,
\eeqa
where $z=z(t)=t\sqrt{1-ib}-i\omega_L/(2\sqrt{1-ib})$  and $w(z)=e^{-z^2} erfc(-iz)$ is the Faddeyeva or $w$-function \cite{FT61,Abra}. Now the amplitude and the phase are defined as $E_{0a}(t)=\sqrt{E^2(t)+\{{\rm{Im}}[E_a(t)]\}^2}$ and $\theta_a(t)={\rm Im}[\ln{E_a(t)}]$, respectively.

{}From Eq. (\ref{E_criterion}) \cite{Cohen}, and for our Gaussian electric field, $E_q(t)\sim E_a(t)$, at least in the central part of the pulse, for   
\beq
\label{Sq_closed_Sa}
\omega_L >> \sqrt{2} \sigma_{\omega q},
\eeq
where $\sigma_{\omega q}=\sqrt{(1^2+b^2)}$ is the bandwidth of $E_q(t)$.

\begin{figure}[h]
\begin{center}
\includegraphics[width=6.cm]{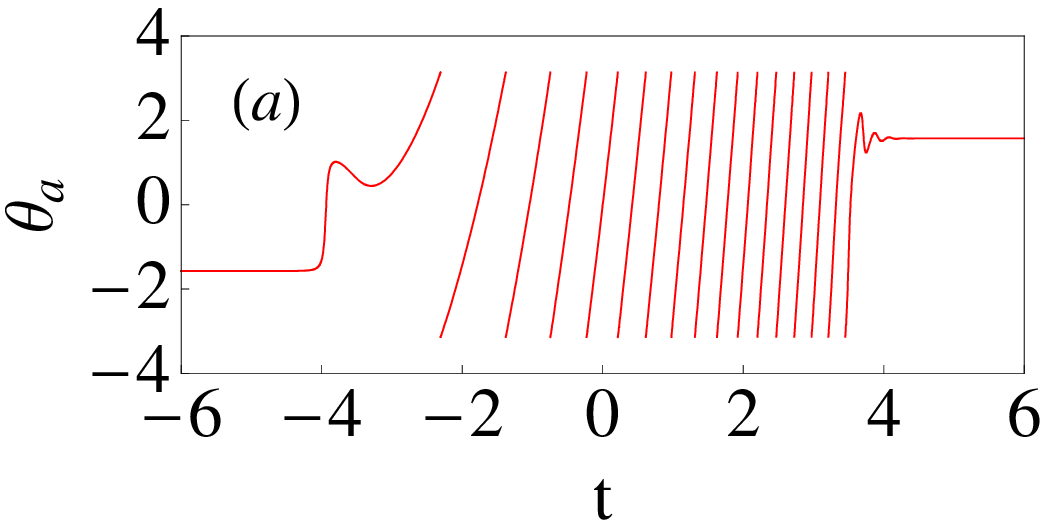}
\includegraphics[width=6.cm]{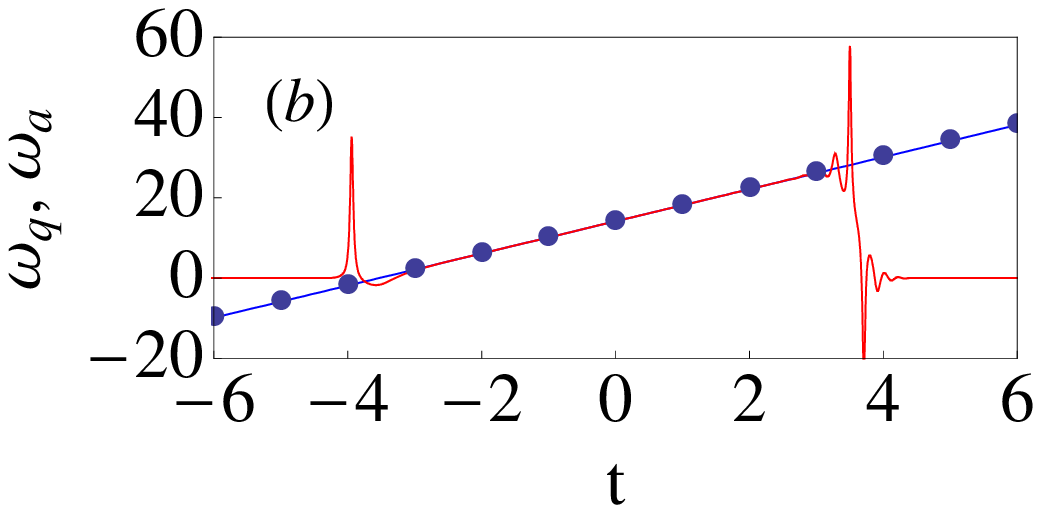}
\caption{(Color online) (a) Phase of the analytic signal $\theta_a(t)$ from $-\pi$ to $\pi$. (b)
The instantaneous frequency $\omega_{a}(t)$ (solid red line) following the linear form of $\omega_{q}(t)=\omega_L+2bt$ (blue line with dots) in the  central interval and tending asymptotically to zero.
Parameters: $b=2$, ${\cal E}_0=4 \sqrt{2}$, and $\omega_L=10 \sqrt{2}$. 
}
\label{phase_frequency}
\end{center}
\end{figure}

We have defined $\theta_q(t)=\omega_Lt+bt^2$ to have a linear chirp, but $\theta_a(t)$ will have in principle a different form. When $E_a(t)\sim E_q(t)$ in a central interval, $\theta_a(t)$ follows the quadratic form of $\theta_q(t)$ there, but it becomes constant
well before and after, see Fig. \ref{phase_frequency}(a). The ``parabola'' is in fact cut into pieces corresponding to the principal branch of the logarithm, from $-\pi$ to $\pi$. Similarly, Fig. \ref{phase_frequency}(b) shows that the instantaneous frequency $\omega_{a}(t)$ follows the linear form of $\omega_{q}(t)=\omega_L+2bt$ in the  central interval but it tends asymptotically to zero. The details of the transition are discussed in the appendix \ref{appendixA}.
\subsection{Population inversion}
If the dynamics is performed exactly, different 
amplitude and phase splittings of the field produce the same results.  Only the real field matters, and the interaction pictures, although different, are all equivalent since they lead to the same Schr\"odinger dynamics and to the same populations. Applying the RWA gives, however, different results.  

Considering that the atom is initially in the ground state, we have solved numerically the system of differential equations for the wave function 
amplitudes with the exact Hamiltonian (\ref{exact_H}), and approximate ones (\ref{HI1_q}) and (\ref{HI1_a}). 

The validity of the RWA is usually discussed for 
time-independent frequencies and 
linked to assuming first the  weak-coupling condition
\cite{Cohen_light, Eberly},
\beq
\label{condition_RWA}
\omega_0 \gg \Omega_R(t),  
\eeq
and then a quasi-resonance condition $|\Delta|/\omega_0\ll 1$. 
This later condition is clearly violated out of resonance 
for ultrachirped pulses. Nevertheless, if the condition (\ref{condition_RWA}) is satisfied and 
$\omega_L$ is also large enough to satisfy the closeness condition (\ref{Sq_closed_Sa}), the populations $P_2(t)$ driven by the two approximate Hamiltonians
are equal to each other and to the population driven by the exact Hamiltonian. 
When (\ref{Sq_closed_Sa}) is not satisfied, however, the populations $P_2(t)$ driven by these approximate Hamiltonians are different and the range of validity of the RWA is much wider for the analytic signal than for the quadrature, as we shall see.

\begin{figure}[h]
\begin{center}
\includegraphics[height=4cm,angle=0]{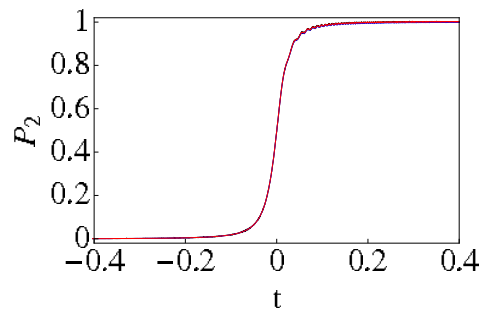}
\end{center}
\caption{\label{P2_2000}
(Color online) Population of the excited state for the exact dynamics  and for the RWA approximations with the quadrature signal and the analytic signal. The carrier frequency is large enough so that the three lines essentially coincide. Parameters: $b=3500$, ${\cal E}_0=200$, and $\omega_L=20000$.}
\end{figure}

\begin{figure}[h]
\begin{center}
\includegraphics[width=6.cm]{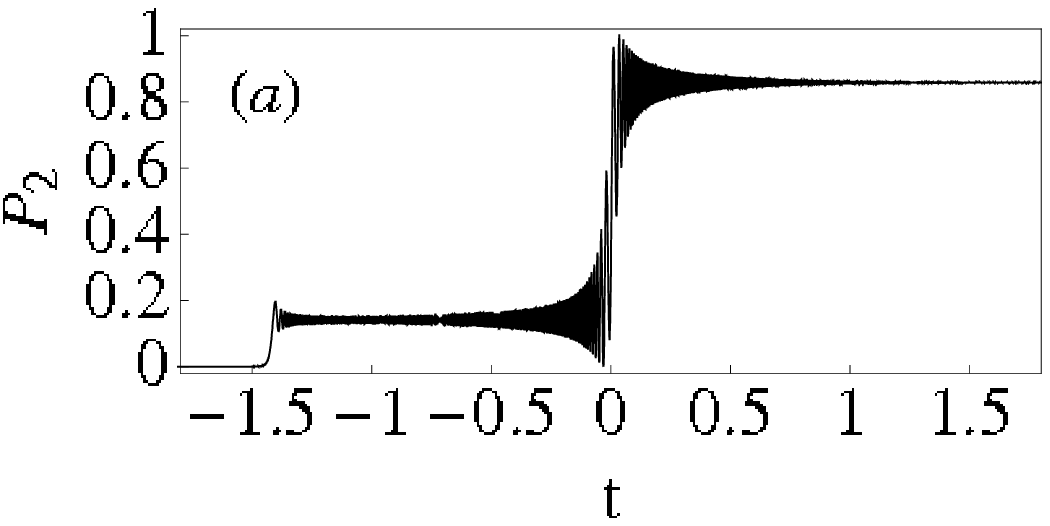}
\includegraphics[width=6.cm]{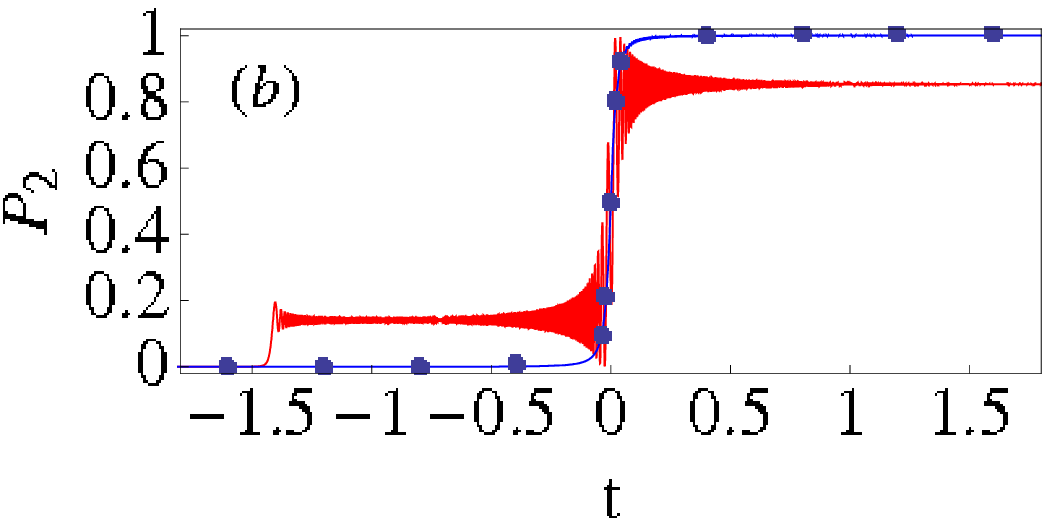}
\caption{(Color online) Population of the excited state for (a) the exact dynamics (b) the Hamiltonian in the RWA for the quadrature signal (blue line with dots) and the analytic signal (solid red line). Compare to the previous figure. As $\omega_L$ decreases, only the analytic signal RWA follows the exact dynamics. Parameters: $b=3500$, ${\cal E}_0=200$, and $\omega_L=5000$.}
\label{P2_500}
\end{center}
\end{figure}

\begin{figure}[h]
\begin{center}
\includegraphics[width=6.cm]{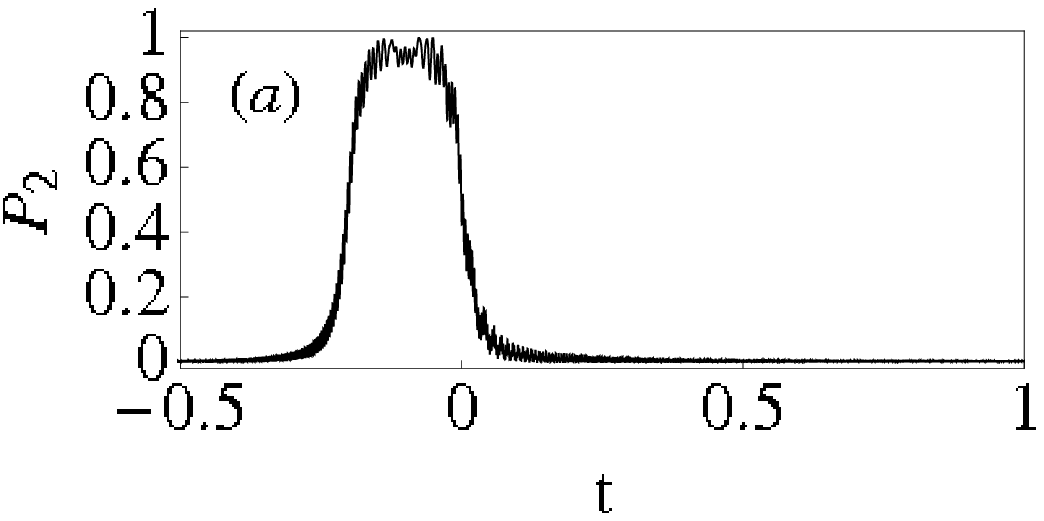}
\includegraphics[width=6.cm]{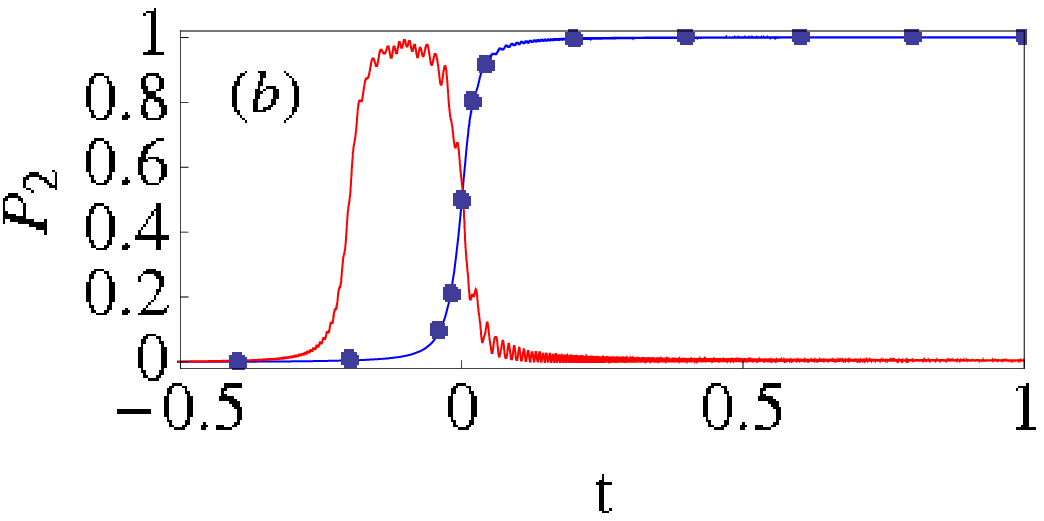}
\caption{(Color online) Population of the excited state for: (a) exact dynamics; (b) RWA for the quadrature signal (blue line with dots) and the analytic signal (solid red line). 
Parameters: $b=3500$, ${\cal E}_0=200$,  $\omega_L=700$.}
\label{P2_70}
\end{center}
\end{figure}

The parameters of the system, considering that $\omega_0(t)=\omega_L$, are ${\cal{E}}_0$, $\omega_L$, and $b$. 
For the values 
${\cal{E}}_0=200$, $b=3500$ and  
$\omega_L=20000$, see Fig. \ref{P2_2000},
the population is fully transferred to the excited state according to the 
exact dynamics. We then decrease $\omega_L$ up to $700$.    
Figures \ref{P2_2000}-\ref{P2_70}
show that in this range of carrier frequencies, the population of the excited state $P_2(t)$ with the analytic signal RWA follows the exact results, while
the quadrature RWA fails. At $\omega_L=700$, see Fig. \ref{P2_70}, the quadrature RWA model 
predicts full inversion, whereas the analytic RWA model and exact results give a 
complete population return.

This behavior shows up that the condition (\ref{condition_RWA}) for the quadrature RWA is much more strict than for the analytic signal RWA. 
To reach the agreement between the quadrature RWA and the exact dynamics we 
have to set $\omega_L=20000$, $100$ times greater than the maximum of $\Omega_{Rq}(t)$, $\Omega_{Rq}(0)=200$, whereas the analytic signal RWA still follows the exact results when $\omega_L=700$, only $3.5$ times greater than the maximum, $\Omega_{Ra}(0)=200.08$. 
For $\omega_L\approx600$ (figure not shown), a small discrepancy 
appears between the populations of the exact and analytic RWA model
which increases with decreasing $\omega_L$.   

The complete population return shown in Fig. \ref{P2_70} is stable, 
for $b$ between $b=2000$ to $b=4000$. Since varying $b$ amounts to 
shifting the NF resonance, this provides ``windows'' of excitation of controllable duration, potentially useful for optical gating, 
with switching times much shorter than the pulse duration.  
   

\begin{figure}[h]
\begin{center}
\includegraphics[height=4cm,angle=0]{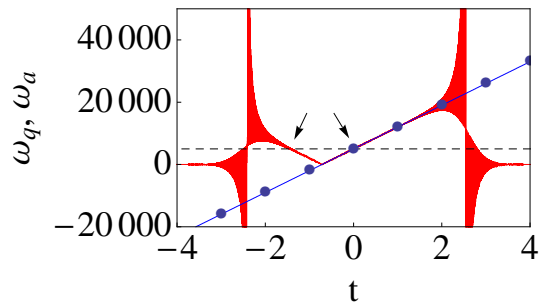}
\end{center}
\caption{\label{frequency}
(Color online) Instantaneous frequencies $\omega_{q}(t)=\omega_L+2bt$ (blue line with dots) and $\omega_{a}(t)$ (solid red line). Parameters: $b=3500$, ${\cal E}_0=200$, and $\omega_L=5000$. The dashed line is at $\omega_L$. The arrows indicate the early, and nominal resonance regions for the pulse, when $\omega_a=\omega_L$.}
\end{figure}
\begin{figure}[h]
\begin{center}
\includegraphics[height=4cm,angle=0]{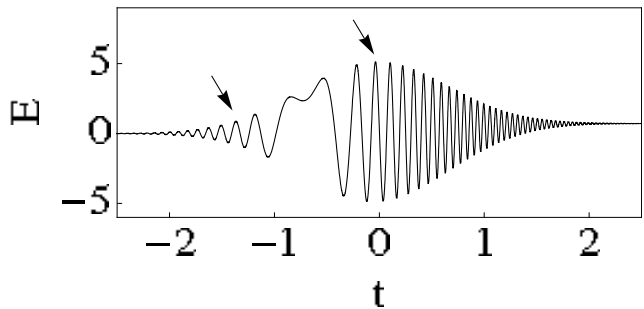}
\end{center}
\caption{\label{30540}
Ultrachirped signal. The arrows indicate the resonance regions where 
$\omega_L=\omega_a$. Parameters: $b=30$, ${\cal E}_0=5$, and $\omega_L=40$.}
\end{figure}

There is, according to the construction of the pulse and the condition $\omega_L=\omega_0$,
a resonance region at $t=0$. 
The early excitation at negative times in Figs.  \ref{P2_500}-\ref{P2_70} 
is  
due to the crossing of an additional resonance region where the instantaneous frequency becomes, in the quadrature model,
$-\omega_0$, see Fig. \ref{frequency}. 
Using as before $\omega_0=\omega_L$, we have that $|\omega_L|= \omega_L+2bt$ 
has solutions $t=0$, corresponding to the nominal resonance of the pulse,
and also 
\beq\label{timer}
t=t_r:=-\omega_L/b.
\eeq
As shown in Figs. \ref{P2_500}-\ref{P2_70}, the quadrature RWA does not
notice this ``negative-frequency'' (NF) resonance, 
since the negative instantaneous frequency
corresponds in fact to a large quadrature detuning $\Delta_q=2 \omega_0$ at $t_r$.  
Instead, the instantaneous frequency for the analytic signal 
is positive at $t_r$ and gives $\Delta_a=0$, see Fig. \ref{frequency}.
The condition for the NF resonance to occur out of the pulse 
(therefore having no consequence) is that $t_r >> 3\sigma_t$, i.e., 
\beq
\label{second_resonance}
\omega_L >> 3b/2.
\eeq
This is independent of the magnitude of the interaction. 
Since the difference between the single-resonance condition (\ref{second_resonance}) and the closeness condition (\ref{Sq_closed_Sa}) is just a numerical factor, when $b>>1$, the separation of analytic and quadrature approximate RWA dynamics is associated with the occurrence of this 
early NF resonance within the pulse. 
Fig. \ref{30540} depicts an example of an ultrachirped real signal capable of inducing a double resonance. The passage through zero instantaneous frequency is evident as a gap in the lower 
envelop and nearby slow oscillations. 
\subsection{Adiabatic approximation}
%
%
%
It is interesting to complement the above results by examining the validity of the adiabatic approximation. 
The condition for the states evolving with (\ref{HI1_q}) and (\ref{HI1_a}) to behave adiabatically following the instantaneous 
eigenstates is
\beqa
\label{adiabaticity_condition}
|\Omega_d(t)| \gg \frac{1}{2}|\Omega_{Ad}(t)|,
\eeqa
where the subscript $d=q,a$ refers to the amplitude-phase decomposition, quadrature or analytic signal, $\Omega_d(t)=\sqrt{\Omega_{Rd}^2+\Delta_d^2}$ is an effective Rabi frequency,  
the instantaneous energies for  the Hamiltonians (\ref{HI1_q}) and (\ref{HI1_a})
are $E_{d\pm}(t)= \Omega_{d}(t)/2$, and 
$\Omega_{Ad}(t)=[\Omega_{Rd} \dot{\Delta}_d(t)-\Delta_d(t) \dot{\Omega}_{Rd}(t)]/\Omega_d^2(t)$.



In our numerical example, according to Eq. (\ref{Sq_closed_Sa}), $\omega_L=20000$ is large enough to make the analytic and the quadrature signal models 
very similar. 
For this $\omega_L$ the adiabaticity condition is satisfied both for $E_q(t)$ 
and for $E_a(t)$. As we diminish $\omega_L$, however, the adiabaticity condition is still satisfied for $E_q$, but not for $E_a(t)$. This is quite evident e.g. in Fig. \ref{P2_500}, note the smooth curve of the quadrature signal model versus the coherent transients of the exact results or the analytic signal model \cite{Chat2001,Chat2006}.

%
%
%
\section{Discussion. Conclusions}
In this paper we have explored the consequences of going beyond 
common excitation regimes of a two-level quantum system,
in particular we have seen that for ultrachirped fields the resonance condition
may be satisfied twice, for positive and negative quadrature instantaneous frequencies along the linear chirp.
The formal rotating wave approximation is more robust by using analytic signal theory for the complex signal and the corresponding interaction picture. The reason is that the analytic signal reinterprets the chirp, assigning positive
frequencies to the pulse region with negative frequencies in the quadrature model. 

Numerical examples demonstrate that the necessary dimensional parameters
to see double resonances are within reach in the microwave domain with the current technology of pulse generators. 
A dimensional realization of the
parameters of Fig. \ref{P2_70} could be as follows: $\omega_L=2\pi\cdot 1.019$ GHz, $a=(2\pi)^2 \cdot0,21\cdot 10^{13}$ Hz$^2$; $b=(2\pi)^2\cdot7.4\cdot10^{15}$ Hz$^2$, which gives $\sigma_t=54.652$ ns, and $\sigma_{\omega q}=2\pi\cdot 5.09$ GHz.  

While this work has been essentially a curiosity-driven exploration, 
applications may be envisioned in state determination, optical gating and interferometry, since the timing of the resonance regions can be controlled.           
Specific fields we consider for future work 
are: Circuit Quantum Electrodynamics, 
in which the two-level system and its interactions are highly
tunable and potentially time dependent, and also cold atoms in counterpropagating laser beams inducing a Raman transition \cite{Cha}.   

\section*{Acknowledgments}
We thank L. Cohen, I. Lizuain, R. Montero, and E. Solano for discussions.
We acknowledge 
funding by the Basque Government (Grant No. IT472-10)
and Ministerio de 
Ciencia e Innovaci\'on (FIS2009-12773-C02-01).
S. I. acknowledges a Basque Government fellowship (Grant No. 
BFI09.39). 
\appendix
\section{Asymptotic values of the phase of the analytic signal}\label{appendixA}
For a linearly chirped Gaussian pulse the phase $\theta_a(t)={\rm{ Im}}[\ln{E_a(t)}]$ of the analytic signal (\ref{Ea_W}) tends 
to a constant, see Fig. \ref{phase_frequency}. 
$z=z(t)=t\sqrt{1-ib}-i\omega_L/(2\sqrt{1-ib})$
is a linear function of $t$, so as $|t|$ increases $z(t)$ becomes larger
in modulus and asymptotic expressions of the $w$ may be used. 
The asymptotic behavior of $E_a(t)$ for large $|z|$ and ${\rm{Im}}[z(t)]>0$ is  
\beqa
\label{Ea_expansion1}
E_a(t) &\sim&
\frac{{\cal E}_0}{2} \bigg[ \frac{ie^{-\omega_L^2/4(1-ib)}}{\sqrt{\pi}z(t)} + \frac{ie^{-\omega_L^2/4(1+ib)}}{\sqrt{\pi}z^*(t)}
\\
&+& 2e^{-\omega_L^2/4(1+ib)} e^{-(z^*)^2(t)}\bigg], 
\eeqa
whereas for ${\rm{Im}}[z(t)]<0$ \cite{W_function}, 
\beqa
\label{Ea_expansion2}
E_a(t)&\sim& \frac{{\cal E}_0}{2} \bigg[\frac{ie^{-\omega_L^2/4(1-ib)}}{\sqrt{\pi}z(t)} + \frac{ie^{-\omega_L^2/4(1+ib)}}{\sqrt{\pi}z^*(t)}
\nonumber\\
&+& 2e^{-\omega_L^2/4(1-ib)} e^{-z^2(t)} \bigg].
\eeqa
%
\begin{figure}[h]
\begin{center}
\includegraphics[height=4.cm,angle=0]{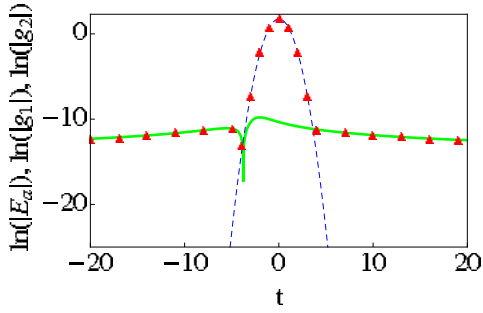}
\end{center}
\caption{\label{logEa}
(Color online) $\ln{(|E_a(t)|)}$ (red triangles) is essentially equal to $\ln{(|g_2(t)|)}$  (dashed blue line) 
in the central interval, and to $\ln{(|g_1(t)|)}$ (green solid line) in the asymptotic regions. Parameters: $b=2$, ${\cal E}_0=4 \sqrt{2}$, and $\omega_L=10 \sqrt{2}$.}
\end{figure}
Then,
\beqa
\label{asymptotic_expansion}
\theta_a(t) \sim \arctan{\left\{ \frac{F(t) \pm 2 e^{-t^2} \sin{[\theta_q(t)]}}{2 e^{-t^2} \cos{[\theta_q(t)]}}\right\} }  + 2\pi n,
\eeqa
where the negative and positive signs correspond to 
${{\rm{Im}}}(z)>0$ and ${{\rm{Im}}}(z)<0$, respectively, 
$n=0,1,2,...$ and
\beqa
F(t)&=& \frac{2(1+b^2)^{1/4}[t\cos{\alpha}+(bt+\frac{\omega_L}{2})\sin{\alpha}]}{\sqrt{\pi}[t^2+(bt+\frac{\omega_L}{2})^2]} 
\nonumber\\
&\times& e^{-\omega_L^2/4(1+b^2)},
\eeqa
with
\beqa
\alpha= \frac{b\omega_L^2}{4(1+b^2)} + \frac{1}{2}\arctan{b}.
\eeqa
%
Independently of the signs $\pm$ in Eq. (\ref{asymptotic_expansion}), when $t\rightarrow\pm\infty$, the argument of the arctangent 
tends to $+ \infty$ or to $-\infty$. Therefore, the asymptotic expansion for the phase of the analytic signal will be
\beqa
\label{asymptotic_expansion_constant}
\theta_a(t)\sim \pm \frac{\pi}{2}+ 2\pi n.
\eeqa
For $n=0$, the phase remains inside the principal branch of the $\ln{E_a(t)}$, see Fig. \ref{phase_frequency}. As the phase reaches these constant values, $\omega_{a}(t)\rightarrow 0$ when $t\rightarrow \pm \infty$.

To identify the transition between the central pulse and asymptotic regimes 
let us compare different terms with the full expression. 
Defining 
\beqa
\label{g1}
g_1(t)= \frac{{\cal{E}}_0}{2} \left[ \frac{ie^{-\omega_L^2/4(1-ib)}}{\sqrt{\pi}z(t)} + \frac{ie^{-\omega_L^2/4(1+ib)}}{\sqrt{\pi}z^*(t)}\right],
\eeqa
\beqa
\label{g2}
g_2(t)=\left\{
\begin{array}{ll}
{\cal{E}}_0 e^{-\omega_L^2/4(1+ib)} e^{-(z^*)^2(t)}& Im[z(t)]>0
\\
{\cal{E}}_0 e^{-\omega_L^2/4(1-ib)} e^{-z^2(t)}& Im[z(t)]<0
\end{array}
\right.
\eeqa
and comparing 
$\ln{[|E_a(t)|]}$ with $\ln{[|g_1(t)|]}$ and $\ln{[|g_2(t)|]}$, we see that 
(\ref{g2}) dominates during the central interval, and (\ref{g1}) in the outer 
time regions, see Fig. \ref{logEa}. Transition regions with interferences 
occur near the times when $\ln{[|g_1(t)|]}$ and $\ln{[|g_2(t)|]}$ are equal. For the parameters $b=2$, ${\cal{E}}_0=4 \sqrt{2}$, and $\omega_L=10 \sqrt{2}$, see Fig. \ref{logEa}, these instants are $t_1=-3.862$ and $t_2=3.596$, 
so the pulse  is essentially within the central interval.





\end{document}